\documentstyle[aps,preprint]{revtex}
\begin{document}
\newcommand{\beq}{\begin{equation}}
\newcommand{\eeq}{\end{equation}}
\newcommand{\bea}{\begin{eqnarray}}
\newcommand{\eea}{\end{eqnarray}}
\newcommand{\ba}{\begin{array}}
\newcommand{\ea}{\end{array}}
\newcommand{\bc}{\begin{center}}
\newcommand{\ec}{\end{center}}
\newcommand{\lsimeq}{\stackrel{<}{\scriptstyle\sim}}
\newcommand{\gsimeq}{\stackrel{>}{\scriptstyle\sim}}
\newcommand{\ie}{{\it i.e.}}
\newcommand{\bml}{\begin{mathletters}}
\newcommand{\eml}{\end{mathletters}}
\newcommand{\half}{\hbox{$1\over2$}}
\newcommand{\fourth}{\hbox{$1\over4$}}
\newcommand{\eight}{\hbox{$1\over8$}}

\title{Driving superfluidity with photoassociation}
\author{Matt Mackie}
\address{Helsinki Institute of Physics, University of Helsinki, PL 9,
FIN-00014 Helsingin yliopisto, Finland}
\author{Eddy Timmermans}
\address{T-4, Los Alamos National Laboratory, Los Alamos, New Mexico
87545, USA}
\author{Robin C\^{o}t\'{e} and Juha Javanainen}
\address{Department of Physics, University of Connecticut, Storrs,
Connecticut 06269-3046, USA}
\maketitle

\begin{abstract}
We theoretically examine photoassociation of a
two-component Fermi degenerate gas. Our focus is on
adjusting the atom-atom interaction, and thereby increaing
the critical temperature of the BCS transition to the superfluid
state. In order to avoid spontaneous decay of the molecules,
the photoassociating light must be far-off resonance. Very 
high light intensities are therefore required for effective
control of the BCS transition.
\end{abstract}

As the field of quantum degenerate vapors emerges from its
burgeoning adolescence, attention is increasingly shifting from
Bose~\cite{BEC} to Fermi~\cite{FDG} systems. Beyond degeneracy
itself, much of this effort is concentrated on observing the
BCS transition to a superfluid state. However, such
investigations are currently at an impasse: the lowest
temperature that has been achieved is about a third of the Fermi
temperature~\cite{PBLOCK}, whereas the formation of Cooper
pairs requires temperatures colder by at least an order of
magnitude~\cite{LiBCS}. Rather than finesse the experiments to
lower the temperature of the gas further, a more fruitful
approach might be to adjust the atom-atom interaction so as to
{\it raise} the value of the critical temperature. Possible
means for adjustment include the  magnetic-field-induced
Feshbach resonance~\cite{FESH}, rf microwave fields~\cite{RF},
dc electric  fields~\cite{DCE}, and
photoassociation~\cite{PA,OPUSMAG}. In fact, the Feshbach
resonance has been recently applied in this manner to induce
condensation in the otherwise incondensable
$^{85}$Rb~\cite{85RB}, and was additionally explored
(theoretically) for its usefulness in spurring the superfluid
transition~\cite{BOHN}.

The purpose of this paper is to investigate the utility of
photoassociation for inducing the BCS transition.
We~\cite{JAV99,JAV99a,OPUSMAG} and others~\cite{DRU98} have
earlier written down field theories for photoassociation of
bosons. In the present case we consider instead a binary mixture
of fermionic atoms, given by the fields
$\phi_\pm({\bf r})$, photoassociating into a bosonic molecule, given by
the field
$\psi({\bf r})$. The fermions would typically be two states with
different $z$ components of angular momentum in the same atom. As a
result of the
Pauli exclusion principle, there is no $s$-wave photoassociation for
two
atoms in the same internal state, but such a restriction does not apply
to
two different spin components.

We thus have a model Hamiltonian density
governing photoassociation,
\bea
{{\cal H}\over\hbar} &=& -\phi_+^\dagger{\hbar\nabla^2\over2m}\phi_+
-\phi_-^\dagger{\hbar\nabla^2\over2m}\phi_-
+\psi^\dagger\left[-{\hbar\nabla^2\over4m}+
\delta-\half i \gamma_s \right]\psi\nonumber\\
&-& \left[ {\cal D}\,\psi^\dagger\phi_+\phi_-+{\cal D}^*
\phi_-^\dagger\phi_+^\dagger\psi\right]+{4\pi\hbar a\over
m}\,\phi^\dagger_-\phi^\dagger_+\phi_+\phi_-\,,
\label{HD}
\eea
where $m$ is the mass of an atom and $\delta$ is the detuning of the
laser
from the threshold of photodissociation. The detuning is
positive when the
photodissociation (inverse of photoassociation) channel is open.
The coupling strength for photoassociation is $\cal{D}\,$. It may be
deduced
implicitly from Refs.~\cite{JAV99,JAV99a}, and is discussed explicitly
in
Ref.~\cite{OPUSMAG}. Either way, we have
\beq
|{\cal D}({\bf r})| = \lim_{v\rightarrow0}\sqrt{\pi\hbar^2\Gamma({\bf
r})\over v\mu^2}\,.
\label{MATELE}
\eeq
Here $\Gamma({\bf r})$ is the photodissociation rate that a light with
the
intensity prevailing at ${\bf r}$ would cause in a nondegenerate gas of
molecules, given that the laser is tuned in such a way that the
relative
velocity of the molecular fragments (atoms) is $v$, and $\mu=m/2$
is
the reduced mass of two atoms. Because of the statistics, there is a
factor
of $\sqrt{2}$ difference in Eq.~(\ref{MATELE}) from the corresponding
expression for identical bosons. Finally, we have included an
interspecies collisional interaction governed by the $s$-wave
scattering
length $a$ in the Hamiltonian.

Under ordinary circumstances photoassociation by absorption of a
photon leads to a molecular state that is unstable against
spontaneous emission. There is no particular reason why
spontaneous break-up of a primarily photoassociated molecule
would deposit the ensuing atoms back to the degenerate
Fermi gases. A spontaneously decaying molecule is considered
lost for our purposes. Correspondingly, we add to the Hamiltonian a
nonhermitian term proportional to the spontaneous emission rate
of the molecular state $\gamma_s$.

The Heisenberg equation of motion for the molecular field $\psi$
is
\beq
i\dot\psi=
\left[-{\hbar\nabla^2\over4m}+ \delta-\half i \gamma_s \right]\psi
-{\cal D}\,\phi_+\phi_-\,.
\label{HEQ}
\eeq
We assume that the detuning $\delta$ is the largest frequency parameter
in
the problem, and solve Eq.~(\ref{HEQ}) adiabatically for the
field $\psi$. In the process we keep the imaginary part in the energy,
and obtain
\beq
\psi\simeq \left[{{\cal D}\over\delta}+ i{\gamma_s{\cal D}\over2\delta^2}
\right]\phi_+\phi_-\,.
\eeq
Inserting into Eq.~(\ref{HD}), we find an effective Hamiltonian
density
for fermions only,
\bea
{{\cal H}\over\hbar} &\simeq&
-\phi_+^\dagger{\hbar\nabla^2\over2m}\phi_+
-\phi_-^\dagger{\hbar\nabla^2\over2m}\phi_- + {4\pi\hbar a\over m}
\,\phi_-^\dagger\phi_+^\dagger\phi_+\phi_-
\nonumber\\
&+& \left[-{|{\cal D}|^2\over\delta}-i\,{\gamma_s|{\cal D}|^2\over2\delta^2}
\right]\phi_-^\dagger\phi_+^\dagger\phi_+\phi_-\,.
\label{HED}
\eea

Let us first ignore the decay term $\propto\gamma_s$.
Equation~(\ref{HED})
displays an added contact interaction between the two spin species, as
if from the
$s$-wave scattering length
\beq
\bar a = -{\frac{{|{\cal D}|^2}\,m }
     {4\,\pi \,\delta \,{{\hbar }}}}\,.
\eeq
The interaction is attractive if the detuning is positive. But an
attractive interaction is exactly what is needed for the BCS
transition. To simplify matters we assume here that the collisional
interaction in the absence of light $\propto a$ is too weak for
experiments on the BCS transition,  and
ignore the native collisions altogether.

The critical temperature for the BCS transition is~\cite{LiBCS}
\beq
T_ c= T_F\exp\left[-{\pi\over2 k_F|\bar a|}\right]=T_F
\exp\left[-{2\pi^2\hbar\delta\over k_Fm|{\cal D}|^2}\right]\,.
\label{TCRIT}
\eeq
Here $k_F = (3\pi^2\rho)^{1/3}$ is the Fermi wave number for the total
density of atoms $\rho$, and $T_F = \hbar^2 k_F^2/2mk_B$
is the corresponding Fermi temperature. Finally, using $(\rho/2)^2$ for
$\phi_-^\dagger\phi_+^\dagger\phi_+\phi_-$, we find the
loss rate per atom due to spontaneous emission from photoassociated
molecules,
\beq
{1\over\tau} = {\gamma_s|{\cal D}|^2\rho\over2\delta^2}\,.
\eeq

To estimate practical experimental numbers, we first note that the
rate of
photoassociation in a nondegenerate sample at temperature $T$
is~\cite{JAV98,MAC99}
\beq
R = \lambda_D^3\rho\, e^{-{\hbar\delta\over k_B T}}\Gamma\equiv \rho
\left({I\over\hbar\omega}\right) \kappa\,.
\label{PARATE}
\eeq
Here $\lambda_D = \sqrt{2 \pi\hbar^2/\mu k_B T}$ is the thermal
deBroglie
wavelength, $I$ is the intensity ($\rm W\,cm^{-2}$) of photoassociating
light, and $\kappa$ ($\rm cm^5$) is the photoassociation rate
coefficients. There may be statistics dependent numerical factors in
Eq.~(\ref{PARATE}). However, in the current literature such factors are
usually ignored, and we write Eq.~(\ref{PARATE}) accordingly.

Using Eq.~(\ref{PARATE}), a calculation or a measurement of the
photoassociation rate in a thermal sample may be converted into a
prediction of effective scattering length, transition temperature, and
lifetime in a degenerate Fermi-Dirac gas. We express the results in
terms of
$\lambdabar=\lambda/2\pi$, wavelength of photoassociating light divided
by
$2\pi$, $\epsilon_R =
\hbar/(2m\lambdabar^2)$, familiar photon recoil frequency, and a
characteristic intensity for the given photoassociation transition,
$I_0$. This gives
\bea
{\bar a\over\lambdabar} &=& 0.0140077\,{I\over I_0}
\,{\epsilon_R\over\delta}\,,\\
{T_c\over T_F} &=& \exp\left[
-36.2478\,{1\over(\lambdabar^3\rho)^{1/3}}\,{\delta\over\epsilon_R}\,{I_0\over
I}
\right]\,,\\
\epsilon_R\tau &=& 4\,{\delta^2\over\epsilon_R\gamma_s}\,
{I_0\over I}\,{1\over \lambdabar^3\rho}\,.
\label{SPEM}
\eea
The obscure numerical factors, powers of 2 and $\pi$, are
there because we want to use the characteristic intensity
for photoassociation defined in Ref.~\cite{OPUSMAG}. For
instance, if the photoassociation rate coefficient $\kappa$ is
known at a temperature $T$ and detuning $\delta$, the critical
intensity is
\beq
I_0 =
 {\sqrt{\pi}\sqrt{\hbar\delta}\,c \hbar^4 \over
2 \kappa m^2(k_BT)^{3/2}\lambdabar^2}\,e^{-\hbar\delta/k_BT}.
\eeq

Detailed microscopic calculations (or measurements) of photoassociation
rates
are sparse, but they exist for the fermionic isotope ${}^6{\rm Li}$ of
lithium~\cite{cote-prl,cote-pra,cote-jms,cote-pri}. Let us consider an
example already discussed in Ref.~\cite{OPUSMAG}, transitions to
the triplet vibrational state $v'=79$ with the binding energy
$1.05\,{\rm cm^{-1}}$. The characteristic intensity is then
$I_0=9.8\,{\rm mW\,cm^{-2}}$, the wavelength is
$\lambda=671\,{\rm nm}$, and  the recoil frequency is
$\epsilon_R = 63.3\times 2\pi\,{\rm kHz}$. We take the decay rate
of
the molecular state to be twice the spontaneous decay rate of the
corresponding atom, so that $\gamma_s = 12\times2\pi\,{\rm MHz}$. In
our
estimate we assume $\lambdabar^3\rho=1$, corresponding to the
density
$\rho=8.21\times10^{14}\,{\rm cm^{-3}}$ that is high but not
unreasonable. It would then take the intermediate detuning $\delta =
2\times2\pi\,10^{14} \,{\rm Hz}$ and the intensity $I=460\,{\rm
MW\,cm^{-2}}$ to make $T_c=0.1\,T_F$ and $\tau=10\,{\rm
s}$.

The intensity came out very high for a continuous-wave laser, so it
seems
that the only potential candidate for experiments is a tightly
focused, powerful $\rm CO_2$ laser. Our formalism, though, is based on
the
assumption that the laser is close to a photoassociating resonance. We
need
to amend the calculations to give meaningful estimates for the
$\rm CO_2$ laser, whose electric field is in practice direct
current compared to the molecular transition frequencies
involved.

To this effect we first note that in an ordinary two-level system one
may
carry out perturbation theory both within the rotating-wave
approximation,
and in the quasistatic limit without the rotating-wave approximation as
well. The result is that the quasistatic results are obtained from the
near-resonance formulas by replacing the detuning with the 
molecular transition frequency,
$\delta\rightarrow\omega_0$, and multiplying the intensity by two,
$I\rightarrow2I$. Applying this substitution to the scattering length,
at
$\lambdabar^3\rho=1$ we find that the intensity required for
$T_c=0.1\,T_F$ again becomes $460\,{\rm MW\,cm^{-2}}$. With the
same substitutions, the lifetime would be about 20~s. However, as the
frequency of the CO${}_2$ laser is $1/16$ of the resonance frequency
for
photoassociation, the phase space for spontaneously emitted photons is
reduced, and the actual rate of spontaneous emission would be reduced
by an
extra factor of at least $16^2\sim300$. It is clear that spontaneous
emission is not an issue with CO${}_2$ laser excitation.

Up to this point we have only considered photoassociation with one
molecular state, the triplet state with vibrational quantum number
$v'=79$. Now, in lithium as well as in other alkali atoms, most of the
transition strength for dipole transitions starting from the ground
state is in the $D$ lines. Just a few electronic states in a molecule
then inherit most of the transition strength for photoassociation. We
only consider the singlet and triplet excited manifolds in the
${}^6{\rm Li}$ dimer, for which calculations of the
photoassociation matrix elements exist for all vibrational
states~\cite{cote-prl,cote-pra,cote-jms,cote-pri}. It turns out
that the triplet state $v'=79$ carries about the factor $0.07$
of the total transition strength for photoassociation of low-energy 
atoms.
As one should obviously add the changes of the scattering lengths due 
to
all molecular states, in our CO${}_2$ laser example the intensity also
gets multiplied by $0.07$ and becomes
$30\,{\rm MW}\,{\rm cm}^{-2}$.

It is, in principle, possible to tailor the scattering length by
off-resonant photoassociation, and thereby effect the BCS transition in
a
low-temperature Fermi gas of, say, ${}^6{\rm Li}$ vapor. The required
laser
intensities, however, are high. As in the case of coherent
photoassociation~\cite{OPUSMAG}, the problem is not so much that
the matrix elements for photoassociation are weak, but that the
primarily photoassociated molecules tend to decay spontaneously
and the sample is lost. To avoid spontaneous emission, one has
to go very far off resonance, which leads to
challenging requirements on laser intensity. In
pursuit of BCS transition by means of off-resonant
photoassociation, it might be worthwhile to try and look for
other ways of getting around the spontaneous emission.


\begin{thebibliography}{}
\bibitem{cote-pri} It should be noted that the photoassociation rates
calculated in Refs.~\protect\cite{cote-prl,cote-pra,cote-jms} are
inadvertently low by a factor of $(2\pi)^5$.
\end{thebibliography}
\end{document}